# The Quantum Mechanics of Being and Its Manifestation[1]

Ulrich Mohrhoff
Sri Aurobindo International Centre of Education
Pondicherry 605002 India

**Abstract**

How can quantum mechanics be (i) the fundamental theoretical framework of contemporary physics and (ii) a probability calculus that presupposes the events to which, and on the basis of which, it assigns probabilities? The question is answered without invoking knowledge or observers, by interpreting the necessary distinction between two kinds of physical quantities — unconditionally definite quantities and quantities that have values only if they are measured — as a distinction between the manifested world and its manifestation.

Quantum mechanics is seen by many as the fundamental theoretical framework of contemporary physics. To the extent that the theory is testable, however, it is but a probability calculus: the so-called quantum state is determined by a preparation of the system (which, as every experimental physicist knows, includes a classical description of the setup), and it assigns probabilities to the possible outcomes of any subsequent measurement. A quantum state thus presupposes not only a classically describable setup but also classically describable outcome-indicating devices with classically describable outcome-indicating properties. Since quantum mechanics presupposes these things, it cannot be called upon to account for their existence. How, then, can it be the fundamental theoretical framework of contemporary physics?

One way to dispose of this problem is to deny its existence, either by asserting that quantum mechanics cannot ultimately be the fundamental theoretical framework for physics or by denying that quantum mechanics is essentially a probability calculus. Among those who haven't yet given up on the challenge to understand what quantum mechanics is trying to tell us about the world, the second option is vastly more popular. It does, however, raise the problem of objectification, and this has been shown to be insoluble (Busch et al., 1996; Mittelstaedt, 1998). What measurement theorists mean by "objectification" is the coming into existence of an actual outcome (as against an entangled state of the system, the apparatus, and possibly the observer) at the end of a measurement process. Insoluble problems are likely to arise from false assumptions. In this particular case the false assumption is that quantum mechanics ought to account for the existence of the events to which it serves to assign probabilities.

Proponents of the many-worlds extravaganza claim to have "solved" this problem by letting the universe, including observers, split into as many copies of itself as there are outcomes every time something qualifying as a measurement takes place. Suffice it to say that many-worlds interpretations, like other realist interpretations of the wave function, face a number of issues that have by no means been resolved (Barrett, 1999; Saunders et al., 2010; Marchildon, 2015). At the root of all such issues is the problem of conjuring correlata out of correlations or events out of probabilities of events.

---

[1] Published (without the Appendix) in *Cosmology* Vol. 24, April 2, 2016: http://cosmology.com/ConsciousnessUniverse3.html.
While the published paper touches on various ways in which quantum mechanics does *not* have to do with consciousness, the Appendix concerns what quantum mechanics *has* to do with consciousness.

Does it help to invoke the consciousness of the observer or to take the view that quantum mechanics is an epistemic theory, concerned not with the world per se but with our knowledge or information about it? What can be rejected at once is the view that nature obeys the unitary laws of quantum mechanics except when an outcome-indicating property "enters" the consciousness of an observer. It is not surprising that von Neumann (1931), the inventor of the tripartite formulation of the process of measurement (consisting of system preparation, unitary evolution, and objectification), felt compelled to hold this view.

Those who hold with Peierls (1991) that a quantum state "represents our *knowledge* of the system we are trying to describe" (original emphasis) are saying two things: a quantum state is a "compendium of probabilities" (Fuchs and Peres, 2000), which is correct, and probabilities are inherently epistemic, which isn't. Until the advent of quantum mechanics all known probabilities were epistemic; they were ignorance probabilities. We resort to such probabilities whenever there are unknown matters of fact — matters of fact that would allow us to make predictions with certainty if they were known. If there are no matters of fact that, if known, would allow us to make predictions with certainty, the reason we cannot do better than assign probabilities is not lack of knowledge. This gives us every right to look upon the probabilities we then assign as objective. (It's just plain unfortunate that the term "objective probability" came to be used for something that isn't a probability, to wit, a relative frequency.)

Here is another reason why (or another sense in which) quantum-mechanical probabilities are objective. The objects of everyday experience "occupy" space (i.e., they have spatial extent), and they neither explode nor collapse as soon as they come into being. If quantum mechanics has anything to say about these objects, it is that they are "made" of finite numbers of particles which lack spatial extent (quarks and electrons), and which are therefore routinely described as pointlike. (Three quarks "make" a nucleon; a finite number of nucleons "make" an atomic nucleus; and a finite number of nuclei and electrons "make" the chair you trust to support you.) Thanks to quantum mechanics, we also know that the existence of space-occupying objects rests on the objective fuzziness of their internal relative positions and momenta (Mohrhoff, 2009a, 2011a). The standard term for this fuzziness — "uncertainty" — is seriously misleading, for what "fluffs out" those objects cannot be anyone's ignorance of the exact values of their internal relative positions and momenta. It can only be an objective indeterminacy of these values. What, then, is the proper way to describe the objective indeterminacy of a physical quantity? It is to assign probabilities to the possible outcomes of a measurement of this quantity. But if we quantify objective indeterminacies by means of probability distributions, the probabilities used for this purpose have every right to be considered objective.

As said, quantum mechanics presupposes outcome-indicating devices with outcome-indicating properties. This means it requires us to distinguish between two kinds of measurable quantities: those that have definite values if and only if they are measured, and those that possess definite values whether or not they are measured. In a two-slit experiment, for instance, the slit taken by a particle has a definite value (left or right) only if it is measured, while the outcome-indicating property, from which the slit taken by the particle can be inferred (no matter whether anyone is around to make the inference), has an unconditionally definite value. What is the meaning of this dualism? What does it tell us about the nature of Nature? Arguably this is the most profound question raised by the quantum theory.

A possible answer is to invoke the age-old metaphysical distinction between the world as we know or experience it and the world as it is in itself, and to argue that unconditionally definite quantities belong to the former while quantities that have values only if they are measured belong to the latter. Nobody has defended this view more persistently and more consistently than d'Espagnat (1989, 1995), who distinguished between an empirical reality and a reality independent of human minds, which is "veiled." Because quantum mechanics forces us to make this distinction, he argued, "the full elision of the subject" (Bitbol, 1990) cannot be achieved. We cannot pretend that quantum mechanics describes a reality independent of human minds.

It is, however, possible to make sense of the necessity of distinguishing between the two kinds of measurable quantities without invoking consciousness, experience, or the subject. The key is to view the distinction as a distinction between the *manifested world* and its *manifestation* (Mohrhoff 2014ab, 2016). This view has the further advantage of extending our knowledge beyond the empirical reality of d'Espagnat, which corresponds to the manifested world. While the correlata belong to the manifested world, the correlations — those between the outcomes of measurements made on the same system at different times as well as those between the outcomes of measurements made on different systems at the same or at different times — extend our knowledge beyond the manifested world. The possibility of knowledge concerning the manifestation of the world, moreover, argues against a reduction of objectivity to inter-subjectivity. If the objective world would correspond to the experienced world (minus the position and the time whence it is experienced by a subject), there could be no such knowledge.

So what does quantum mechanics tell us about the manifestation of the world? Let us begin by considering the following scattering experiment. Initially two identical particles — particles lacking properties by which they can be distinguished — are found moving northward and southward, respectively. The next thing we know is that the same two particles are found to be moving eastward and westward, respectively. The question then is: which incoming particle is identical with which outgoing particle? It is well known that this question has no answer. The distinction we make between the two possible identifications cannot be objectified (i.e., cannot be regarded as corresponding to an objective difference).

Here as elsewhere, unanswerable questions tend to arise from false assumptions. In this particular case, the question implicitly assumes that we are dealing with *two* things rather than with the same thing detected twice — a single entity initially moving *both* northward and southward and subsequently moving *both* eastward and westward. If the incoming particles (and therefore the outgoing ones as well) are one and the same entity, the question "Which is which?" can no longer be asked. What needs to be borne in mind here is that quantum mechanics does not tell us what (if anything) happens between measurements, except other measurements. As Peres (1984) put it succinctly, "there is no interpolating wave function giving the 'state of the system' between measurements." What's more, there is no compelling reason to believe that the intrinsic identity of the two particles ceases when it ceases to have observable consequences owing to the presence of properties by which they can be distinguished and re-indentified. We are free to take the view that all particles in existence are identical in the strong sense of numerical identity. What presents itself here and now with these properties and what presents itself there and then with those properties is one and the same entity. I shall refer to it simply as "Being."

While fundamental particles are routinely described as pointlike, what is meant is that they lack internal structure. Lack of internal structure is consistent with either a pointlike form or no form at all. See Mohrhoff (2014a, Sect. 9) for reasons why fundamental particles ought to be conceived as formless. But if every fundamental particle in existence is (i) identically the same Being and (ii) formless, then the shapes of things resolve themselves into reflexive spatial relations — i.e., relations between Being and Being. By entering into (or entertaining) reflexive spatial relations (relative positions and relative orientations), Being supports (i) what looks like a multiplicity of relata if the reflexive quality of the relations is ignored, and (ii) what looks like a substantial expanse if the spatial quality of the relations is reified.

This way of thinking goes farther in relationism — the doctrine that space and time are a family of spatial and temporal relations holding among the material constituents of the universe — in that it affirms that the "ultimate material constituents" are (i) formless and (ii) numerically identical. It also demolishes the notion that the physical world can be understood in terms of (a multitude of) ultimate constituents and of the ways they interact and combine. The manifestation of the world is essentially the manifestation of material *forms*. Instead of being constituents of material *things* and parts of the manifested world, subatomic particles, atoms, and molecules are instrumental in the manifestation of material forms. They occupy a (conceptual) position intermediate between Being and the manifested world.

Because the manifestation of the world includes the manifestation of space and time, it cannot be conceived as a process that takes place in time. We keep looking for the origin of the universe at the beginning of time, but this is an error of perspective. The origin of the universe is Being, and the manifestation of the universe is an atemporal transition from undifferentiated Being to a world that is maximally differentiated spacewise as well as timewise. Maximally but not completely, for the manifested world is not differentiated "all the way down" (Mohrhoff, 2009a, Sect. 7; 2011a, Sect. 10; 2014a, Sect. 4).

Here, in brief, is why. A detector is needed not only to indicate the presence of a particle in a region of space but also — and in the first place — to *realize* or *define* a region, so as to make it possible to attribute to a particle the property of being inside. Speaking more generally, a macroscopic apparatus is needed not only to indicate the possession of a property by a quantum system but also — and in the first place — to make a set of properties available for attribution to the system. In addition, macroscopic clocks are needed to realize attributable times. This, of course, is vintage Bohr (1935), who rightly insisted that the "procedure of measurement has an essential influence on the conditions on which the very definition of the physical quantities in question rests." But if detectors are needed to realize regions of space, space cannot be intrinsically partitioned. It is partitioned only to the extent that the requisite detectors are physically possible. Because this extent is limited by the "uncertainty" principle, physical space cannot be realistically modeled as an actually existing manifold of intrinsically distinct points. In other words, the spatial differentiation of the physical world is incomplete. And because macroscopic clocks are needed to realize attributable times, a similar argument leads to the conclusion that the temporal differentiation of the physical world is incomplete as well. Quantum theory thus reverses the explanatory arrow of both common sense and classical physics. Instead of allowing us to explain wholes in terms of their interacting parts, it suggests to us how the multiplicity of the world emerges from an intrinsically undifferentiated Being.

The transition from the unqualified unity of Being to the multiplicity of the macroworld passes through several stages. Across these stages, the world's differentiation into distinguishable regions of space and distinguishable objects with definite properties is being gradually realized. There is a stage at which Being presents itself as a multitude of formless particles. This stage is probed by high-energy physics and known to us through correlations between the counterfactual clicks of imagined detectors, i.e., in terms of transition probabilities between in-states and out-states. There are stages that mark the emergence of form, albeit a type of form that cannot yet be visualized. The forms of nucleons, nuclei, and atoms can only be mathematically described, as probability distributions over abstract spaces of increasingly higher dimensions. At energies low enough for atoms to be stable, it becomes possible to conceive of objects with fixed numbers of components, and these we describe in terms of correlations between the possible outcomes of unperformed measurements. The next stage, closest to the manifested world, contains the first objects with forms that can be visualized — the atomic configurations of molecules — but it is only the final stage — the manifested, macroscopic world — that contains the actual detector clicks and the actual measurement outcomes that allow us to test the correlations that quantum mechanics predicts.

Many of the mysteries surrounding quantum mechanics become clear in this light. Why, after all, is the general theoretical framework of contemporary physics a probability calculus, and why are its probabilities assigned to measurement outcomes? If quantum mechanics concerns a transition through which the differentiation of reality into distinguishable objects and distinguishable regions of space is gradually realized, the question arises as to how the intermediate stages are to be described — the stages at which the differentiation is incomplete and the distinguishability between objects or regions of space is only partially realized. The answer to this question is that whatever is not completely distinguishable can only be described by assigning probabilities to what is completely distinguishable, namely, to the different possible outcomes of a measurement. What is instrumental in the manifestation of the world can only be described in terms of what happens in the manifested world, or else in terms of correlations between events that could happen in the manifested world. (Think of the textbook descriptions of the stationary states of a hydrogen atom, which are correlations between preparations — measurements determining the atom's energy, its total angular momentum, and a component of its angular momentum — and probability distributions assigned on the basis of the outcomes of these measurements.)

But is it even consistent with quantum mechanics to regard certain measurable quantities as definite *per se*? Here is a related question: are there localizable particles? According to a theorem due to Clifton and Halvorson (2002), there is no quantum state such that the probability of finding a particle in a finite region of space is 1. From this, Clifton and Halvorson have drawn the conclusion that the experience of detecting particles in finite regions of space is "illusory" and "strictly fictional." What they have actually shown is that particles cannot be localized relative to the spacetime manifold ***M*** postulated by quantum field theory. But ***M*** is not where experiments are performed. What is illusory is the notion that attributable positions are defined by spatial regions of ***M***. Attributable positions are defined by the sensitive regions of detectors, which, according to said theorem, also cannot be localized in any finite region of space. What is strictly fictional therefore is ***M***, inasmuch as this cannot be localized relative to the positions that particles can possess.

The positions of detectors, in turn, are defined by the positions of macroscopic objects (macroscopic positions, for short). "Macroscopic" is one of the most elusive terms routinely used by physicists. What makes it possible at last to rigorously define it is that the spatial differentiation of the world doesn't go "all the way down" (Mohrhoff 2009a, 2014a). Here is the argument in brief: In a world that is incompletely differentiated spacewise, the next best thing to an object with a sharp position is an object whose position probability distribution is and remains so narrow that there are no detectors with narrower position probability distributions — detectors that could probe the region over which the object's position extends. The events by which the values of macroscopic positions are indicated are therefore correlated in ways that are consistent with the laws of motion that quantum mechanics yields in the classical limit. (There is one necessary exception: in order to permit a macroscopic object — the proverbial pointer — to indicate a measured value, its position must be allowed to change unpredictably if and when it serves to indicate a measured value.) What makes it possible to treat macroscopic positions as definite *per se* is that macroscopic objects follow trajectories that are only counterfactually indefinite. Their positions are "smeared out" only in relation to an imaginary spatiotemporal background that is more differentiated than the manifested world. In a word, macroscopic objects follow definite trajectories because they *define* what we mean by a (definite) trajectory, and they have persistent identities because they follow (definite) trajectories.

## Appendix

What holds the key to the mysterious presence of consciousness in what appears to be a material universe is the self-identical Being that constitutes every particle in existence. The root of consciousness is not to be found in the manifested world, nor in the process of manifestation, but in that which manifests the world. For Being does not simply manifest the world (by entering into reflexive spatial relations); Being manifests the world *to itself*. Being relates to the world not only as the substance that constitutes it but also as the consciousness that contains it. It is at once the single substance by which the world exists and the ultimate self or subject for which it exists.

How, then, are we as conscious beings related to this ultimate self or subject? This question has been answered in considerable detail and on a solid experiential foundation by the Indian philosopher (and freedom fighter, and mystic) Sri Aurobindo (Heehs, 2008). In keeping with a more than millennium-long philosophical tradition (Phillips, 1995), Sri Aurobindo (2005) posits an Ultimate Reality whose intrinsic nature is (objectively speaking) infinite Quality and (subjectively speaking) infinite Delight. This has the power to manifest its inherent

Quality/Delight in finite forms, and the closest description of this manifestation is that of an all-powerful consciousness creating its own content.

In the native poise of this consciousness, its single self is coextensive with its content and identical with the substance that constitutes the content. There, but only there, it is true that *esse est percipi* (to be is to be perceived).

A first self-modification of this *supramental* consciousness leads to a poise in which the one self adopts a multitude of standpoints, localizing itself multiply within the content of its consciousness and viewing the same in perspective. It is in this secondary poise that the dimensions of experiential space (viewer-centered depth and lateral extent) come into being. It is also here that the dichotomy between subject and object, or self and substance, becomes a reality.

Probably the most adequate description of the process by which the one original self assumes a multitude of standpoints is that of a multiple concentration of consciousness. A further self-modification of the original creative consciousness occurs when this multiple concentration becomes exclusive. We all know the phenomenon of exclusive concentration, when consciousness is focused on a single object or task, while other goings-on are registered subconsciously, if at all. A similar phenomenon transforms individuals who are conscious of their essential identity into individuals who have lost sight of this identity and, as a consequence, have lost access to the supramental view of things. Their consciousness is mental, which means not only that it belongs to what appears to be a separate individual but also that it perceives or presents the world as a multitude of separate objects. Mentally conscious beings thus come into existence not *only* by an evolution from seemingly unconscious matter but *also*, and in the first place, by a multiple exclusive concentration of the creative consciousness inherent in Being.

If this multiple exclusive concentration is carried to its logical conclusion, the result is a world whose inhabitants lack both the ability to generate ideas (which is a function of the principle of mind) and the power to execute them (which is a function of the principle of life). And since the latter is also responsible for the existence of individual forms, the result is a world of formless individuals — the fundamental particles of physics. This is how (from our temporal perspective) the original creative consciousness came to be "involved" in mind, how mind came to be "involved" in life, and how life came to be "involved" in formless particles. (If the form of a material object resolves itself into its internal spatial relations, a fundamental particle, lacking internal relations, is a formless entity.) And because these principles are involved in formless particles, matter is capable of evolving life, life is capable of evolving mind, and mental consciousness can and eventually will evolve the supramental consciousness — the power by which Being manifests the world.

How does Being manifest a cosmos (rather than pure, indeterministic chaos) in which life and mind (and supermind) are "involved"? Clearly, Being's reflexive spatial relations must be governed by seemingly inflexible laws possibly of a statistical nature. Furthermore, setting the stage for the drama of evolution calls for objects that are spatially extended (they "occupy" space) and are sufficiently stable (they neither explode nor collapse as soon as they are formed).

Because that stage has been set by carrying the multiple exclusive concentration of consciousness to its logical conclusion, such objects will be, or will appear to be, "made of" finite numbers of formless particles — particles that do not "occupy" space. As I have argued elsewhere (Mohrhoff, 2002, 2009b, 2011b), the existence of such objects not only implies the validity of quantum mechanics but also goes a long way toward establishing the other well-tested laws of contemporary physics (the Standard Model and General Relativity). These laws, then, are preconditions of the possibility of an evolutionary manifestation of Being — a Being that relates to the world not only as the substance by which it exists but also as a (supramental) consciousness for which it exists.

## References

Barrett, J. A. (1999). The quantum mechanics of minds and worlds (Oxford University Press, Oxford).

Bitbol, M. (1990). L' Elision. Preface to Schrödinger, E., L'esprit et la matière (Seuil, Paris).

Bohr, N. (1935). Quantum mechanics and physical reality. Nature 136, 65.

Busch, P., Lahti, P. J., and Mittelstaedt, P. (1996). The quantum theory of measurement, 2nd revised edition, Sect. III.6.2 (Springer, Berlin).

Clifton, R., and Halvorson, H. (2002). No place for particles in relativistic quantum theories, Philos. Sci. 69, 1–28.

D'Espagnat, B. (1989). Reality and the physicist (Cambridge University Press, Cambridge, UK).

D'Espagnat, B. (1995). Veiled Reality (Addison-Wesley, Reading, MA).

Fuchs, C. A., and Peres, A. (2000). Quantum theory needs no 'interpretation.' Phys. Today 53 (3), 70–71.

Heehs, P. (2008). The lives of Sri Aurobindo (Columbia University Press, New York).

Marchildon, L. (2015). Multiplicity in Everett's interpretation of quantum mechanics. Stud. Hist. Philos. M. P. 52B, 274–284.

Mittelstaedt, P. (1998). The interpretation of quantum mechanics and the measurement process, Sect. 4.3b. (Cambridge University Press, Cambridge, UK).

Mohrhoff, U. (2002). Why the laws of physics are just so. Found. Phys. 32 (8), 1313–1324.

Mohrhoff, U. (2009a). Objective probability and quantum fuzziness. Found. Phys. 39 (2), 137–155.

Mohrhoff, U. (2009b). Quantum mechanics explained. Int. J. Quantum Inf. 7 (1), 435–458.

Mohrhoff, U. (2011a). A fuzzy world. In Vision of oneness, edited by Licata, I., and Sakaji, A. J. (Aracne editrice, Ariccia, Italy), pp. 41–61.

Mohrhoff, U. (2011b). The world according to quantum mechanics: why the laws of physics make perfect sense after all, Chap. 22 (World Scientific Publishing, Singapore).

Mohrhoff, U. (2014a). Manifesting the quantum world. Found. Phys. 44 (6), 641–677.

Mohrhoff, U. (2014b). Quantum mechanics and the manifestation of the world. Quantum Stud.: Math. Found. 1 (3–4), 195–202.

Mohrhoff, U. (2016). Quantum mechanics in a new light. Found. Sci. DOI 10.1007/s10699-016-9487-6.

Peierls, R. (1991). In defence of 'measurement.' Phys. World 4 (1), 19–20.

Peres, A. (1984). What is a state vector? Am. J. Phys. 52, 644–650.

Phillips, S. H. (1995). Classical Indian metaphysics (Open Court, Chicago/La Salle).

Saunders, S., Barrett, J., Kent, A., and Wallace, D. (2010). Many worlds? Everett, quantum theory, and reality. (Oxford University Press, Oxford).

Sri Aurobindo (2005). The life divine (Sri Aurobindo Ashram Publication Department, Pondicherry, India).

Von Neumann, J. (1931). Mathematische Grundlagen der Quantenmechanik (Springer, Berlin). English translation: Mathematical foundations of quantum mechanics (Princeton University Press, Princeton, 1955).